# Study of strong photon–magnon coupling in a YIG–film split–ring resonant system

B. Bhoi[2*], T. Cliff[1*], I. S. Maksymov[1], M. Kostylev[1†], R. Aiyar[2], N. Venkataramani[3], S. Prasad[2], and R. L. Stamps[1,4]

[1]*School of Physics M013, University of Western Australia, Crawley 6009, Australia*
[2]*Department of Physics, Indian Institute of Technology Bombay, Powai, Mumbai 400076, India*
[3]*Department of Metallurgical Engineering and Materials Science, Indian Institute of Technology Bombay, Mumbai 400076, India*
[4]*SUPA, University of Glasgow, Glasgow, G12 8QQ, United Kingdom*

**Abstract:** By using the stripline Microwave Vector–Network Analyser Ferromagnetic Resonance and Pulsed Inductive Microwave Magnetometry spectroscopy techniques, we study a strong coupling regime of magnons to microwave photons in the planar geometry of a lithographically formed split–ring resonator (SRR) loaded by a single–crystal epitaxial yttrium–iron–garnet (YIG) film. Strong anti-crossing of the photon modes of SRR and of the magnon modes of the YIG film is observed in the applied-magnetic-field resolved measurements. The coupling strength extracted from the experimental data reaches 9% at 3 GHz.

Theoretically, we propose an equivalent circuit model of the SRR loaded by a magnetic film. This model follows from the results of our numerical simulations of the microwave field structure of the SRR and of the magnetisation dynamics in the YIG film driven by the microwave currents in the SRR. The equivalent-circuit model is in good agreement with the experiment. It provides simple physical explanation of the process of mode anti-crossing.

Our findings are important for future applications in microwave quantum photonic devices as well as in nonlinear and magnetically tuneable metamaterials exploiting the strong coupling of magnons to microwave photons.

## 1. Introduction

In order to be useful for quantum application, a proposed technology has to be able to exchange information with preserved coherence [1-3]. To this end, a system which consists of two sub–systems has to operate in a regime called 'strong coupling'. The strong coupling regime is characterised by strength of coupling between the subsystems which is larger than the mean energy loss in both of them. A straightforward way to decrease losses is to make use of resonant systems [4–6]. Systems that exploit plasmonic resonances make use of strong coupling between electric dipoles and optical fields localised at the sub-wavelength scale (i.e. on a spatial scale smaller than the wavelength of light). This makes possible the creation of solid–state sources of quantum states of light (single photons, indistinguishable or entangled photons) such as those based on semiconductor quantum dots (QDs) or nitrogen–vacancy centres in diamonds (NVs), embedded in optical microcavities [7–10] or precisely placed in close proximity to an optical nanoantenna [11–13].

---
[*] Both authors contributed equally to this paper.
[†] Corresponding author

On the other hand, strong light–matter coupling is also achievable by using magnetic dipoles [14]. These schemes make possible the processing of quantum information in various microwave resonator systems comprising ultra–cold atomic clouds [15], molecules [16], NVs [17, 18] and ion–doped crystals [19].

Recently, strong coupling of microwave photons to magnons was demonstrated for a system representing a microwave cavity loaded by a single–crystal yttrium–iron–garnet (YIG) sphere [1, 20–22]. Furthermore, a more planar geometry of a microwave stripline resonator loaded by a single–crystal YIG film was shown to produce strength of coupling of 150 MHz [23].

Nowadays, a microstrip line has largely replaced a traditional microwave cavity as a source of driving microwave magnetic field in the ferromagnetic resonance (FMR) experiment [24, 25]. On the other hand, the stripline arrangement is used to investigate planar microwave metamaterials [26]. Consequently, it makes this technique very suitable for the investigation of coupling of microwave photons to magnons. Additionally, the use of microstrip lines opens up avenues for the development of magnetically tuneable metamaterials based on arrays of highly–conductive meta–molecules (resonant elements that comprise metamaterials) coupled to magnetically–active materials of different sizes and shapes [27, 28].

In a recent work – Ref [23], the interaction between a non–magnetic split–ring resonator (SRR) and a thin film of YIG was investigated with the Microwave Vector–Network Analyser (VNA) Ferromagnetic Resonance (FMR) method. Strongly hybridised resonances were observed. The YIG film was grown on a single–crystal yttrium aluminium garnet substrate by pulsed laser deposition using an excimer laser. Although the SRR supported an anti–symmetric (low–frequency) and a symmetric (high–frequency) mode, only strong coupling of the YIG static magnetic field–dependent mode to the symmetric mode was studied, and a coupling strength of 150 MHz was observed.

In the present work, we study the interaction of magnetic resonances in a YIG film with microwave photon resonances in SRRs. We use both the VNA–FMR (frequency domain) and Pulsed Inductive Microwave Magnetometry ('PIMM', time domain) spectroscopy techniques to investigate the coupling between photons and magnons. We observe a strong coupling between the YIG mode and both low–frequency anti–symmetric and high–frequency symmetric modes of the SRR. In the VNA–FMR experiment, this interaction manifests itself as a strong anti–crossing between the photon and magnon mode. Naturally, the same result is confirmed by Fourier–transforming time–domain PIMM traces into the frequency domain. Additionally, the inspection of time–domain traces reveals the presence of a strong beat effect. The presence of the beat signal in time–domain traces is a signature of the entanglement of qubits [29, 30].

We conduct numerical simulations of the microwave field structure of the split rings and of the magnetisation dynamics driven by the microwave currents in the ring. We also suggest an equivalent circuit of an SRR loaded by a magnetic film. These calculations are in very good agreement with the experiment and deliver a clear physical picture of the process of anti–crossing of the SRR and magnon modes.

## 2. Experimental arrangement

The investigated SRR (Fig. 1) represents a small metallic loop with a slit in it. It is a kind of *LC* resonant contour in which the loop acts as an inductance and the slit represents a lumped capacitance. Due to the lumped–element approach an SRR can support resonance wavelengths noticeably larger than the linear size of the SRR. An epitaxial single–crystal YIG film also represents a microwave resonant system which supports FMR. FMR is the collective precession of spins about the equilibrium spin direction in the material. In a ferro– or ferrimagnetic materials the spins are coupled by the exchange and dipole–dipole interactions. The energies of the (inhomogeneous) exchange and dipole–dipole interactions determine the FMR frequency. For the present work it is important that the FMR frequency also depends on the static magnetic field [31] in which the material is placed. The dependence on the applied field is through the Zeeman energy contribution to the energy of magnons which represent quanta of FMR [32].

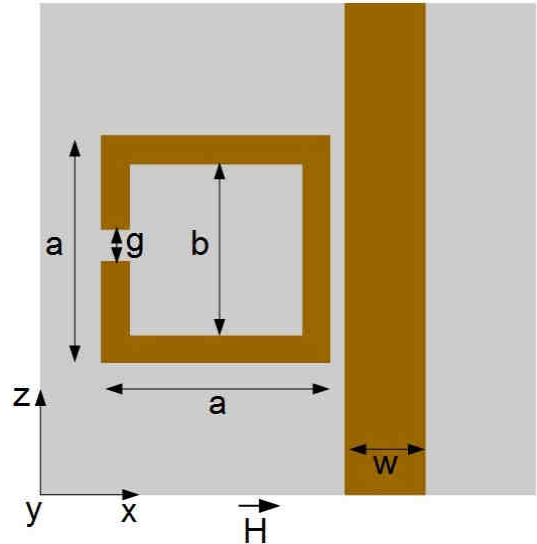

Fig. 1. Sketch of the split ring resonator structure. The split ring is inductively coupled to a microstrip feeding line. In experiment, the input and output of the microstrip line are connected to a VNA and the static applied magnetic field **H** is created by an electromagnet (not shown). For the measurements, a YIG film (not shown) with the dimensions 10 mm × 15 mm × 25 μm is placed on top of the split ring. The measured dimensions of the split ring and the microstrip line are: $a$ = 8.5 mm, $b$ = 7.5 mm, $g$ = 0.06 mm (the distance between the microstrip line and the SRR is also $g$), $w$ = 3 mm. Also not shown: the total thickness of the SRR plus the dielectric substrate (grey area, $\varepsilon$ = 2.3) and the back–side metallisation is $h$ = 0.84 mm. The thickness of the copper lines of the SRR and the microstrip is 0.01 mm [33].

Usually, FMR in a sample is excited by placing the sample in a uniform (or quasi–uniform) microwave magnetic field. An onset of FMR is easily seen as applied–field dependent resonance absorption of the microwave power by the magnetic material. A convenient way to excite FMR in ferro– and ferrimagnetic *films* is by placing them on top of a microstrip or coplanar microwave stripline transmission line [24, 25, 34] or forming a microstrip line directly on top of the film [35]. A microwave current flowing through the stripline induces a microwave Oersted field in the space above the stripline. This field drives spin precession in the material.

A split–ring resonator can be formed lithographically as a loop of a microstrip line on top of a microwave substrate. One can use the Oersted field of a microwave current flowing through such an SRR to drive FMR in a ferrimagnetic film sitting on top of the SRR. In this way, we realise a simple and effective means of implementing additional functionality into the split ring design. This is the central point of our paper.

The geometry of the device under test is shown in Fig. 1. It represents an SRR inductively coupled to a microwave transmission line. A single–crystal yttrium–iron–garnet film grown on a 1 mm–thick gadolinium–gallium garnet (GGG) substrate sits on top of the resonator with the YIG layer facing the SRR. The film is 25 μm thick and was grown with liquid–phase epitaxy (LPE). A static magnetic field **H** is applied in the plane of the film in the direction perpendicular to the microstrip line (Fig. 1).

The measurements have been taken at room temperature. In order to take the measurements in the frequency domain, the input and the output of the microstrip line have been connected to the ports of a VNA and the transmission characteristic of the microstrip line (S21=Re(S21)+$i$Im(S21)) has been measured as a function of microwave frequency $f$ and the strength $H$ of the applied field.

In the alternative time–domain (PIMM) arrangement, the input port of the microstrip line is connected to the output port of a generator of short pulses. We excite our system with a pulse of a rectangular shape which is 1 V in amplitude and 10 ns in duration. The nominal pulse rise time at the output port of the pulse generator is 55 ps.

## 3. Experimental results

A set of representative |S21| vs. microwave frequency $f$ dependencies taken with VNA–FMR for a number of values of the applied field is shown in Fig. 2. In each of the panels (except one for 300 Oe) one observes two peaks. One peak is very strongly dependent on the applied field. Essentially it moves across the displayed frequency range with an increase in the frequency. For $H$=300 Oe this peak has an almost vanishing amplitude and is located at about 2.25 GHz. Its relative amplitude becomes much larger when it moves closer to second – higher–frequency – peak (the peaks at 2.7 GHz and 3.4 GHz respectively in the panel or $H$=430Oe). For $H$=450 Oe the lower–frequency peak ceases moving across, but the higher–frequency peak starts to move quickly with $H$. Its amplitude drops with an increase in the frequency separation from the lower–frequency peak. The respective frequency vs. field dependence of the peak positions are shown in Fig. 3(a). One clearly sees strong anti–crossing of the two lines which suggests strong coupling of the modes.

In order to identify the resonances these measurements have been repeated when the YIG film covered not only the SRR but also a section of the feeding microstrip line. These measurements have been taken on a different SRR structure whose geometry was able to accommodate the film in this position. The shape of the SRR for this structure was the same as in Fig. 1, but its sizes were slightly different, therefore the frequencies of the resonances in Fig. 3(b) do not coincide with the ones in Fig. 3(a) (and in Fig. 4 which we will discuss later on). From Fig. 3(b) one sees that for this YIG film placement one observes a third mode located in between two peaks of the type shown in Fig. 2. The frequency vs. applied field dependence for this extra mode is well fitted by the Kittel formula [36]

$$f = \frac{\gamma}{2\pi}\mu_0\sqrt{H(H+M_s)} \tag{1}$$

for the FMR frequency of an in–plane magnetised film with a saturation magnetisation value $\mu_0 M_s$=0.2 T=2 kOe. (Here $\gamma/(2\pi)$ is gyromagnetic ratio; typically 2.8 MHz/Oe for YIG.) This value is very close to the standard value of $\mu_0 M_s$ for YIG: 0.175 T.

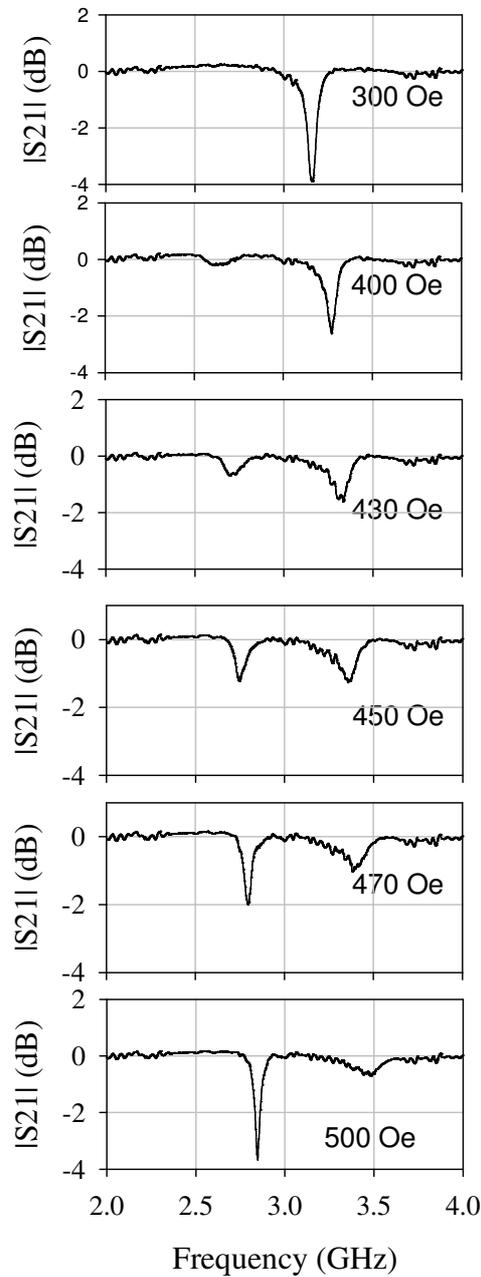

Fig. 2. Representative |S21| vs. microwave frequency traces taken with the vector network analyser. The numbers in the panels are the strengths of the applied static magnetic fields applied to the YIG film to take the measurements.

This suggests that this extra mode is the ferromagnetic resonance mode ('magnon mode') of the YIG film excited directly by the Oersted field of the feeding microstrip line. Because of this origin, it is decoupled from the other two modes. For this reason its frequency vs. applied field dependence is a smooth monotonic function. From Fig. 3(b) one also sees that the four other modes converge with this extra line either at higher or lower frequencies. These four modes clearly separate into two pairs. Each of the pairs contains a mode located on the lower–field side from the magnon mode and a mode located at the higher–field one. Importantly, far away from the "anti–crossing" with the dispersion line the frequencies are almost the same within each pair and the line slope is practically vanishing. This identifies the horizontal sections of these four lines as uncoupled SRR resonances ('photon modes'). The sections of these lines with significant slopes (close to the anti–crossing area) are SRR resonances coupled to magnon modes of the YIG film.

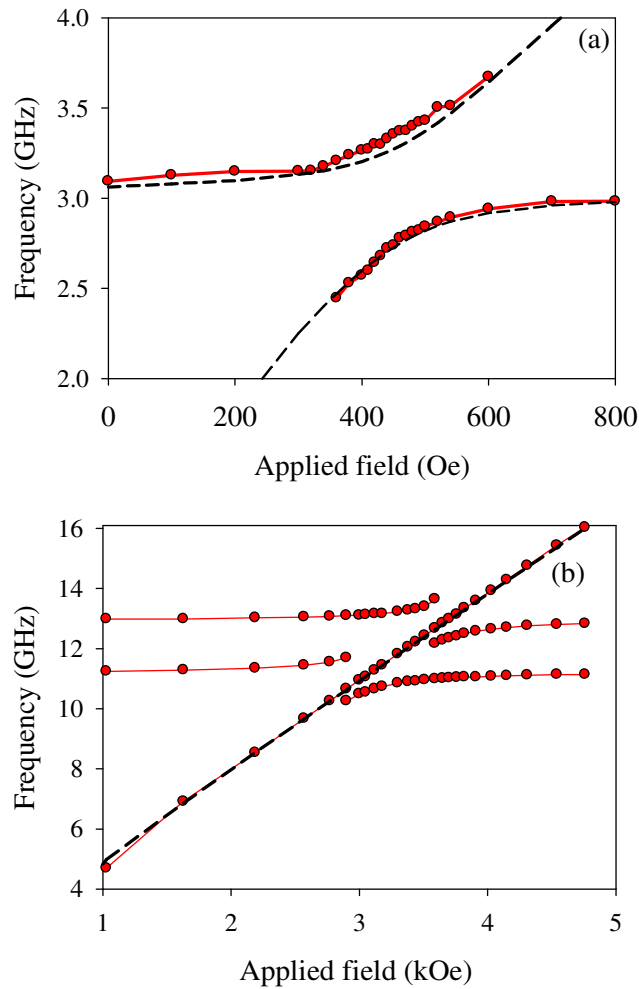

Fig. 3. (a) Frequencies of the peaks from Fig. 1 as functions of the applied field. Dots – experiment, dashed lines – fits with Eq. (2). (b) Results of a measurement when a YIG film covers both an SRR and a section of the microstrip line. Dots – experiment, dashed line – fit with the Kittel formula (1) for FMR frequency of an in–plane magnetised film. The solid lines in both panels are the guides for the eye.

The strong anti–crossing between the photon and magnon modes seen in Fig. 3 suggests a strong coupling between them. Figure 4 displays the results we obtained on the sample from Fig. 1 in a broad range of frequencies and applied fields. Two SRR modes are visible in it: at 3.2 GHz and 6.6 GHz. [the lower one is the same as in Fig. 3(a)]. Both modes strongly interact with the YIG magnon mode.

The dashed lines in Fig. 3(a) are the fits of the experimental data for the lower-frequency section of Fig. 4 with the model of two coupled resonators (see, e.g., Ref. [22])

$$f_{1(2)} = \frac{f_1^0 + f_2^0}{2} \pm \sqrt{\left(\frac{f_1^0 - f_2^0}{2}\right)^2 + \Delta^2} \,, \tag{2}$$

where $f_1$ and $f_2$ are the frequencies of the coupled resonances, $f_1^0$ and $f_2^0$ are the respective resonance frequencies in the absence of coupling and $\Delta$ is the coupling strength (measured in frequency units). By assuming that the frequency $f_2^0$ is given by the Kittel formula Eq. (1) (depends on the applied field) and $f_1^0$ is independent on the frequency (uncoupled SRR mode), from the fit we obtain $\Delta$ = 270 MHz or $\Delta/f_1^0 = 9\%$. Similarly, for the higher–frequency anti–crossing (located between 5 and 7 GHz) the best fit with Eq. (1) is obtained for $\Delta$ = 450MHz or 6.8%.

This latter value is significantly larger than the one previously observed for a pulse–laser deposited YIG film (1.3%) but weaker than the one for a bulk YIG crystal (approximately 20%) [23]. The SRR geometries and sizes in Fig. 1 and in Ref. [23] are essentially different, therefore it is difficult to estimate the contribution of the SRR design to the performance of our device–prototype. However, from the comparison of the three results it becomes certain that the usage of a much thicker YIG film in the present work (25 µm) than in Ref.[23] (2.4 µm), combined with potentially smaller intrinsic magnetic losses for the film in our case delivers a significant contribution to the improvement of the device performance observed in the present work. Indeed, it is known that coupling of magnon dynamics in ferromagnetic films to microwave fields of striplines scales with the film thickness [37] and that the amplitude of any resonance driven by an external source is inversely proportional to the intrinsic losses in the resonating medium. Although the magnetic loss parameter for the YIG film from [23] is not specified in the paper, it is natural to suppose that the film from [23] is characterised by an at least half an order of magnitude larger loss parameter than our YIG film, because the former was grown with pulse laser deposition and the latter with liquid–phase epitaxy (LPE). It is known that LPE is the only method able to produce extremely low–loss films [38]. The strength of coupling of 20% observed in [23] for the case of the bulk YIG crystal is explained in a similar way. It is due to the very large thickness of this material potentially combined with very low magnetic losses, if the used YIG slab represents a single–crystal material.

In Ref. [23] a strong positive peak of |S21| was observed between the two usual negative peaks of resonance absorption for the maximum of anti–crossing of the two resonances (middle line in Fig. 3 in Ref. [23]). It was claimed that this peak could be explained as due to a negative refraction index for the YIG+SRR meta–material in these conditions. Interestingly, we do not observe this behaviour in our Fig. 2. Instead, for *H*=450 Oe we see strong reduction in the peak amplitudes. It reaches 3dB with respect to the peak shown in the panel for 300 Oe. (Note that our simulations in Sect. 3 do not reproduce this extra peak either.)

The result of the time–domain measurements is shown in Fig. 5(a). The main observation in this figure is the damped high–frequency oscillations on top of the rectangular pulse. The shape and the amplitude of this oscillatory pattern strongly depend on the applied field. The results of the Fourier analysis of these time sequences are consistent with the frequency–resolved measurements in Fig. 4 and Fig. 2. In particular, for 430 Oe one observes strong reduction in the amplitude of the high–frequency oscillation and noticeable change in the regularity of oscillations. Also, for 900 Oe and 1300 Oe the dominant period of the temporal sequence is practically the same and practically equal to the one for 0 Oe. This fact can be explained as the dominant contribution to the total oscillation amplitude originating from the lowest SRR resonance (at 3 GHz for this field). This is expected since the Fourier component of the frequency spectrum of the rectangular excitation pulse of duration $\tau$ is given by the simple expression $F(f)=\sin(\pi f\tau)/(\pi f\tau)$. For $\tau = 10$ ns, $F(3.0\ \text{GHz})/F(6.6\ \text{GHz}) = 15$. Hence, the oscillations at 6.6 GHz may be excited by a rectangular pulse less efficiently than at 3.0 GHz by an order of magnitude or so. For all these reasons, the change in the oscillation pattern between 0 Oe and 430 Oe is very noticeable and may be explained as the beat of two damped resonance modes from the respective panel in Fig. 2. On the contrary, the respective beat pattern is practically invisible in the time trace for 1400 Oe. This is because the dominating contribution to this pattern is delivered by the SRR resonance at 3 GHz. However, the Fourier analysis in Fig. 5(b) reveals the presence of this beat pattern.

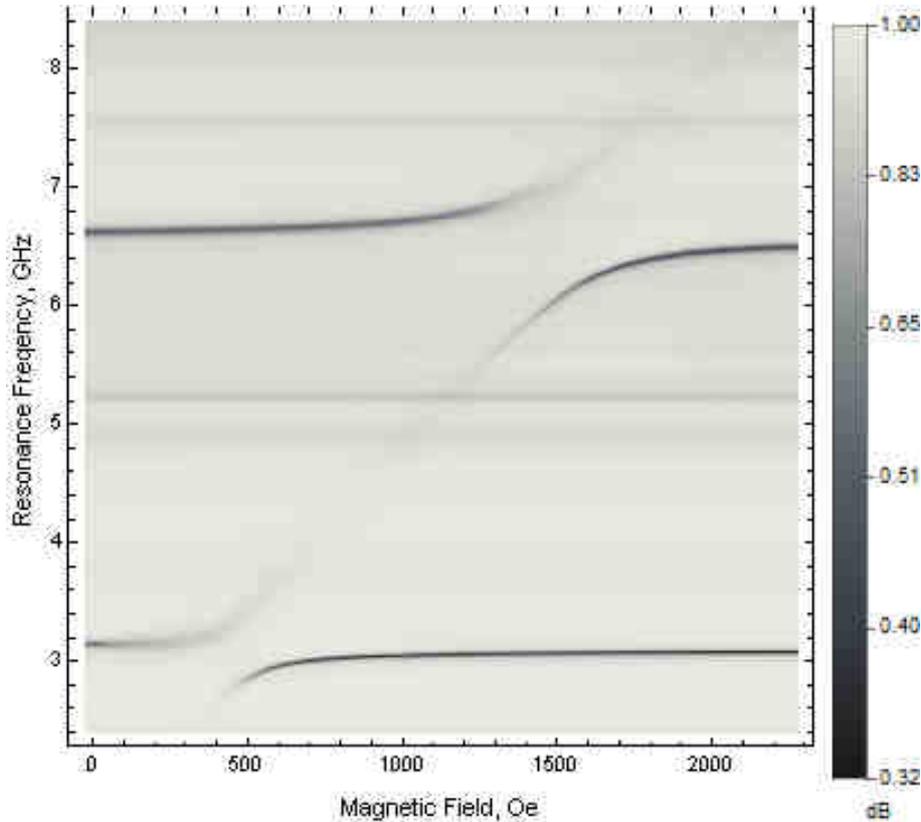

Fig. 4. Grey–scale plot of the linear magnitude of the real part of S21 (Re(S21)) as a function of the applied field and microwave frequency obtained using VNA–FMR.

## 3. Numerical results and discussion

We have conducted rigorous numerical three–dimensional finite–difference time–domain (FDTD) simulations in order to understand electrodynamic properties of the SRR. The FDTD method is a very well–known time–domain Maxwell's equations solver, which allows simulating of open–space problems using so–called absorbing boundary conditions [39]. Conceptually, this numerical method is a counterpart of the time–domain PIMM technique. In brief, we simulate the propagation of a short pulse of microwave current through the microstrip line coupled to the SRR. For the sake of clarity, in the first approximation we consider the SRR without the YIG film. This simulation is repeated for an isolated microstrip line (the SRR is absent). Similar to PIMM, the simulated time–domain traces are Fourier–transformed to obtain transmission characteristics. The transmission characteristic of the isolated microstrip line is used for the normalisation, which removes a standing wave pattern originating from artificial reflections at the ends of the microstrip line, which are unavoidable in our numerical model.

The numerical results are shown in Fig. 6. We see that the FDTD simulation qualitatively reproduces the experimental picture. The low– and the high–frequency resonances of the SRR are easily identified [Fig. 6(a)]. The inspection of the magnetic field profiles of these resonances [Fig. 6(b–e)] reveals that the microwave magnetic field of SRR is strongly localised in close proximity to the split ring; it drops sharply with distance from the ring edges. From this point of view the double–SRR structure from Ref. [23] may be more advantageous since it ensures more complete filling of the area inside the SRR with the microwave magnetic field. Figure 6(f) also confirms the result in Ref. [23] that the profile of the microwave magnetic field for the low–frequency mode is anti–symmetric but the one for the high–frequency mode is symmetric with respect to an imaginary horizontal axis running through the middle of the ring gap in Fig. 1. In the following we will call these SRR modes as 'symmetric' and 'anti–symmetric', respectively.

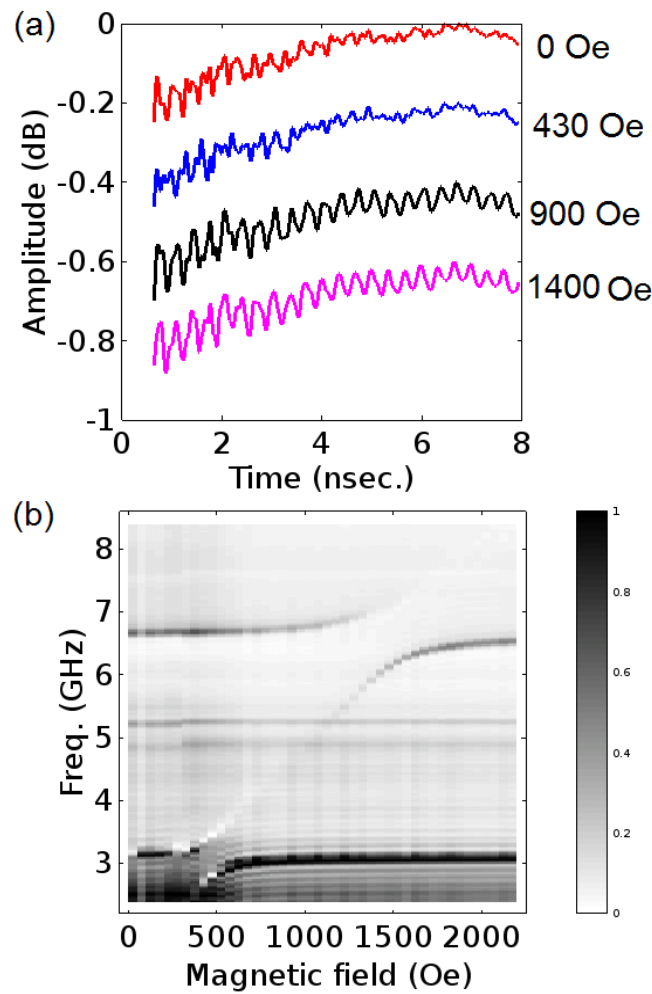

Fig. 5 (a) Time–domain traces for the selected values of the applied field. For the sake of clarity, each trace has been offset vertically by –0.2 dB. (b) Grey–scale plot of the linear magnitude of Fourier–transformed PIMM traces as a function of the applied field and microwave frequency. Note a different grey–scale bar as compared with Fig. 4. We used the 'Window filtering procedure' to remove noise from the time traces and thus improve the quality of the Fourier–space data.

In our VNA–FMR experiment we noticed that the resonant frequencies shift when we remove the YIG film from the SRR. The YIG film is kept on top of a significantly thicker GGG substrate. Due to computation constraints, in our simulations the GGG substrate cannot be as thick as it is in the experiment. Nevertheless, simulations with a significantly thinner GGG substrate (not shown) qualitatively reproduce the experimental results: the resonance frequencies of the SRR are shifted and an additional very strong peak arises at ~6 GHz [40]. This peak is probably due to the microwave resonance of the GGG substrate, and it seems to be realistic as confirmed by the data in Figs. (4) and 5(a). In these figures one sees a weak resonance at about 5.2 GHz which does not interact with the magnon mode at all. The unrealistically large amplitude of this peak in our simulation seems to be an artefact, possibly due to constraints of the numerical model (which assumes much smaller GGG thickness, etc.).

A separate numerical simulation was run to qualitatively explain the origin of the mode anti–crossing. It was based on a completely different approach. As has been mentioned in Ref. [23], the magnetic dynamics excited by microwave magnetic field of the SRR is actually not a genuine ferromagnetic resonance, i.e. the dynamics does not take the form of magnetisation precession whose amplitude is (quasi) uniform and the phase is constant over the film area covered by the SRR. The excitations driven by the microwave field of the SRR are actually travelling spin waves. Two types of spin waves are relevant for our geometry. Both exist in in–plane magnetised ferromagnetic films [41]. The Damon–Eshbach (DE) wave propagates in the film plane at a right angle to the applied magnetic field **H**. It is characterised by a positive dispersion (group velocity) – $d\omega/dk > 0$, where $k$ is the wave number. The backward volume magnetostatic spin wave (BVMSW) is characterised by a negative dispersion $d\omega/dk < 0$. Its wave vector is parallel to the applied field. The frequency (energy) gaps $\omega(k=0)$ for both types of waves are the same and are given by the Kittel formula Eq. (1).

Microwave Oersted fields of microstrip lines are able to efficiently excite spin waves in the wave number range

$$-2\pi/s < k < 2\pi/s, \tag{3}$$

where $s$ is the width of the microstrip (in Fig. 1 $s=(a-b)/2 = 0.5$ mm). The excited wave propagates perpendicular to the longitudinal axis of the microstrip. Hence, given the direction of the applied field (along $x$ in Fig. 1), the two SRR sides parallel to the $x$–axis excite DE waves and the sides along the $z$–axis excite BVMSW. Hereafter, we will focus on the fundamental SRR resonance (at $f_1^0 = 3.106$ GHz in the experiment and at 3.7 GHz in the simulation data in Fig. 6 obtained for the SSR without the YIG film). An important peculiarity of BVMSW excitation is that this type of spin wave is excited by the out–of–plane component ($h_y$) of the microwave magnetic field of the microstrip (see, e.g., Ref. [42]). From Figs. 6(a) and 6(g) one sees that for the fundamental mode this field is practically absent for the SRR side containing the gap; it is present only at the opposite SRR side. Furthermore, the BVMSW excitation is significantly less efficient than that of the DE wave [42], because the latter is excited by both in–plane ($h_z$) and out–of–plane components of the microstrip field. The in–plane microwave field component gives a significantly larger contribution to the total dynamic magnetisation amplitude. The $h_z$–contribution is larger because of elliptical polarisation of spin waves (i.e. ellipticity of magnetisation precession) in in–plane magnetised films [42]. For those reasons, to keep our theory simple, we will neglect the contribution of BVMSW excitation to the total SRR response, and concentrate on the waves which deliver the main contribution to the magnon mode – the DE waves. (If necessary, the BVMSW contribution can be included by using the theory from [42]).

A parameter which is often used to describe the efficiency of spin wave excitation by microstrip transducers is the radiation impedance (see, e.g., Ref. [43]). We use the theory from Ref. [44] to calculate the radiation resistance of spin waves excited by the microwave current in the split–ring resonator. Figure 7(a) shows the calculated dependence of the radiation impedance for a transducer which represents two parallel microstrip lines at a distance 8 mm from each other. The length of the lines is assumed to be equal to *a* in Fig. 1 and the microwave currents in the two parallel microstrips were assumed to be in phase. This arrangement mimics the two SRR sides which are parallel to the *x*–axis. The in–phase currents ensure that the $h_z$ fields for the two sides are in anti–phase as seen in Fig. 6(f) for the mode at 3.7 GHz. The applied field $H$ = 485 Oe was chosen such that the maximum of Re($Z_r$) is at the same frequency as the experimental uncoupled SRR resonance – 3.106 GHz, i.e. $f_2^0(H) = f_1^0$ in the notations of Eq. (1). As a proxy to the uncoupled resonance we consider the case $H$ = 0. For this field value, DE waves exist in the frequency range 0 to 2.8 GHz, but should be excited in a much smaller frequency range, due to the very large value of *s* [see Eq. (3)]. Hence, no magnetisation dynamics with noticeable amplitude are expected for $H$ = 0 at 3.106 GHz.

As seen from Fig. 7(a), the radiation impedance has non–negligible values in a narrow range of frequencies $\delta f$ = 500MHz or so. Analysis of this frequency range demonstrates that it corresponds to the DE wave–number range given by Eq. (3). For *s* = 0.5 mm, $2\pi/s = k_{max} = 125$ cm$^{-1}$. On one hand, this very small value of $k_{max}$ shows that the system operates in the regime which is close to FMR conditions (FMR corresponds to $k$ = 0). On the other hand $\delta f >> (\gamma \Delta H)/(2\pi)$, where $\Delta H$=0.5 Oe, is the standard magnetic loss parameter for LPE YIG films. The latter parameter determines the intrinsic resonance linewidth for FMR. (Basically, $(\gamma \Delta H)/(2\pi)$ is the frequency–resolved linewidth and $\Delta H$ is the field–resolved one). The fulfilment of the condition $\delta f >> (\gamma \Delta H)/(2\pi)$ formally implies that the travelling–spin–wave contribution to the FMR linewidth is significant [45]. This happens because *s* is significantly smaller that the free propagation path for spin–waves in YIG films. (For our 25 μm–thick LPE-grown film the latter amounts to several millimetres.)

Indeed, in Fig. 8(a) one clearly sees travelling spin waves propagating away from the two microstrip lines, placed at a distance *a* from each other. The existence of a travelling spin wave is evidenced by a linear spatial dependence of the amplitude of dynamic magnetisation seen in this figure for |*z*| > 5 mm or so, i.e. outside the SRR perimeter. This linear dependence on the logarithmic scale corresponds to an exponential decay of the dynamic magnetisation amplitude $m_y$ on the linear scale. The exponential decay of amplitude is characteristic to a travelling wave in any medium with non–negligible losses.

One notices a large difference in the wave amplitudes propagating in the positive and the negative direction of the *z*–axis: $m_y(z<-5\text{mm}) > m_y(z>+5\text{mm})$. This is due to the strong non–reciprocity of the DE wave excitation by microstrip transducers (see, e.g., Ref. [42]).

Because the free propagation path of spin waves in YIG is very large, the partial wave excited by the right–hand–side microstrip in Fig. 8(a) reaches the left–hand–side microstrip. This leads to interference of the two partial waves excited by the two parallel sides of the SRR. The interference forms an oscillatory amplitude pattern characteristic to a partial standing wave in the space between the two microstrips (e.g., inside the SRR). This standing–wave pattern is very pronounced in Fig. 8(a) for $-4.25$ mm $= -a/2 < z < +a/2$. A similar oscillatory pattern is seen in Fig. 8(a) for the radiation impedance and is formed because the DE wave number varies with frequency according to the dispersion law for the DE wave [41]. As a result, the phase $\phi = ka$ accumulated by the wave after crossing the distance $a$ between the two sides of the SRR is a function of frequency. The maxima and the minima of the 'interference pattern' in Fig. 7(a) correspond to $\phi$ equal to either an even or odd multiple of $2\pi$.

Figure 8(b) demonstrates an equivalent circuit for the fundamental mode of SRR resonance loaded by the magnon mode in YIG. Because of the much smaller length of the SRR perimeter with the respect to the wavelength of the electromagnetic wave in the SRR substrate, one can consider the SRR as a series *LC* contour with lumped *L* and *C*. Physically, the capacitance *C* is concentrated in the gap. Although the inductance *L* is actually distributed along the perimeter of the SRR we can also consider it as a lumped one for the same reason of the smallness of the perimeter with respect to the microwave wavelength at 3 GHz. The losses in this contour are due to ohmic losses in the SRR's metal; in our model they are accounted for by an equivalent resistance $R_0$. $Z_r$ is an equivalent lumped element whose complex impedance is given by the plots in Fig. 7(a).

The dashed line in Fig. 7(b) shows the reactance $X_0$ of this contour for $H = 0$. As stated above, for this field $Z_r = 0$, hence

$$X_0 = -1/(\omega C) + \omega L, \tag{4}$$

where $\omega = 2\pi f$. The resonance frequency $f_1^0$ of the unloaded SRR follows from the condition $X_0 = 0$. It is given by the well–known formula

$$f_1^0 = 1/\sqrt{LC}. \tag{5}$$

$X_0 = 0$ corresponds to a maximum of the real part of the complex conductance $Y_0$

$$Y_0 = 1/(R_0 + iX_0). \tag{6}$$

In Fig. 7(b) one clearly sees a sharp resonance peak of Re($Y_0$) at $f_1^0 = 3.06$GHz.

For $H = 485$ Oe one has to add $Z_r$ in series to $R_0 + iX_0$. From Fig. 7(a) one sees that there are two frequencies for which Im($Z_r$) $= -X_0$. The existence of the two compensation points is possible because the SRR resonance represents an 'anti–resonance' (given by a maximum of the complex *conductance*) [46], but the spin wave excitation represents a 'resonance' (given by a maximum of the complex *impedance*).

The resonance condition for the whole sequence of the equivalent elements connected in series follows from $X_0 + $Im($Z_r$) $= 0$ or, equivalently, it is given by the frequency position of the maximum of the real part of the total complex conductance

$$Y = 1/(R_0 + iX_0 + Z_r). \tag{7}$$

Accordingly, the presence of the two frequencies for which Im($Z_r$) $= -X_0$ results in two resonances for the coupled system, seen as the anti–crossing of $f_1$ and $f_2$ [in the notations of Eq. (1)] in the field and frequency resolved data in Figs. 3 and 4.

The two peaks for $H=485$ Oe in Fig. 7(b) are the result of calculation using Eq. (7). In this calculation we assume that the value of $C$ is given. Then the value of $L$ is obtained from Eq. (5) by setting $f_1^0$ to the frequency corresponding to the experimental SRR frequency for $H = 0$ (Step 1). The extracted value of $L$ allows us to determine $R_0$ as

$$R_0 = \omega L/Q, \tag{8}$$

where $Q$ is the experimental quality factor for the SRR resonance for $H = 0$ (Step 2). The last step of the calculation (Step 3) is to determine the two frequencies for which $X_0+\text{Im}(Z_r) = 0$. An additional step of the calculation is converting the results for the 'internal' dynamics of the SRR into the signal from the output of the feeding microstrip line. To include the coupling of the loaded SRR to the feeding microstrip line into our model, we introduce an equivalent lumped mutual inductance $M$ into the equivalent circuit in Fig. 8(b). It can be then shown that

$$S21 = 1 - KY, \tag{9}$$

where $K$ is a coefficient – 'SRR to microstrip coupling coefficient' – which depends on $M$.

To obtain the traces of Re($Y$) shown in Fig. 7(b) we fit the experimental data with Eqs. (4–8) using $C$ as a single fit parameter. To this end, we iteratively perform Steps 1–3 for different values of $C$ until the frequency difference between the positions of the two peaks, shown by the solid line in Fig. 7(b), becomes equal to the experimental one from the panel for 470 Oe in Fig. 2: 592 MHz. The $Z_r(f)$ profile is kept the same throughout the fitting procedure; we use the one shown in Fig. 7(a). We choose the particular value of the experimental field – 470 Oe – because from Fig 3(a) one sees that for this field value the frequency separation of the two coupled resonances is a minimum. According to Eq. (1) this implies that for this field $f_2^0(H) = f_1^0 = 3.106$ GHz and hence the results of the calculation in Fig. 7(a) correspond to this particular experimental field value.

As an initial (guess) value for $C$ we employ a value which we extracted from the FDTD simulation of the microwave electric field inside the gap: 4.9 pF. Once the value of $C$ has been extracted from the fit, the value of $K$ may be found by equating Re(S21) in Eq. (9) to the maximum of the negative peak of the experimentally measured Re(S21) for $H = 0$ [Fig. 7(d)].

The best fit is obtained for $C=2.0$ pF which is of the same order of magnitude as the above mentioned guess value. The corresponding values of $L$ and $R_0$ are 1.01 nH and 0.71 Ohm respectively. We believe that both quantities are quite reasonable. The respective plots of Re(S21) are shown in Fig. 7(c). These data correspond to $K = 0.23$.

The main observation from Fig. 7(c) is in good qualitative agreement with the experimental data in Fig. 7(d). Furthermore, in both Fig. 7(c) and 7(d) one sees a fine structure (small amplitude oscillations) on top of the higher–frequency peak. Comparison of Figs. 7(a) and 7(c) reveals that this fine structure is due to the spin–wave interference (see above).

One more observation is that the linewidth of the lower–frequency coupled resonance is smaller than that of the uncoupled one, both in experiment and theory. The decrease in the peak linewidths, due to coupling to the magnon system, may be explained based on the observation that the slope of the Re($Z_r$) vs. $f$ dependence, near the lower–frequency compensation point, is of the same sign as of $X_0(f)$. Therefore Im$\{Z_r(f)\}+X_0(f)$ is a steeper function of $f$ than $X_0(f)$. This leads to quicker detuning from the resonance with $f$ than for the uncoupled resonance and hence a quicker drop in the amplitude with the frequency.

This idea is in agreement with the much larger experimental and theoretical resonance linewidths for the higher–frequency coupled resonance [Figs. 7(c) and 7(d)]. Indeed, if one neglects the fast oscillations of Im($Z_r$) [Fig. 7(a)] and follows the local mean value of Im{$Z_r(f)$}, then one finds that at the frequencies slightly below this resonance Im{$Z_r(f)$} first becomes flatter and then the slope changes from positive to negative. This change in the slope results in a long tail of the higher–frequency coupled resonance peak, spanning over the range of 400 MHz towards smaller frequencies. This tail is very visible in Figs. 7(c) and 7(d); it significantly broadens the upper–frequency peak with respect to the lower-frequency one.

Two small discrepancies between the theory and the experiment become clear from the comparison of Fig. 7(c) and 7(d). The first one is a noticeable shift upwards in frequency of the theoretical resonance pair for $H = 485$ Oe with respect to $H = 0$. Indeed, the experimental pair is located more symmetrically with respect to the $H = 0$ data than the theoretical one. We believe that this may be due to the fact that in our calculation we neglected the BVMSW contribution to $Z_r$. Inclusion of the excitation of BVMSW would modify the lower frequency slope of the peak of Re($Z_r$) by making it less steep. Accordingly, the lower–frequency slope of Im($Z_r$) would become less steep and non–vanishing values of Im($Z_r$) would extend to smaller frequencies. This would potentially move the lower–frequency peak for $H = 485$ Oe in Fig. 7(c) to lower frequencies, thus making the location of the two peaks more symmetric with respect to the $H = 0$ peak.

The other noticeable discrepancy is the peak amplitudes. The experimental data in Fig. 7(d) demonstrate a larger reduction in the peak amplitudes for $H = 470$ Oe with respect to the $H = 0$ case than the theoretical ones. This difference may suggest that $K$ is dependent on $H$. Indeed, the YIG film covers the SRR but does not cover the feeding microstrip line. Therefore one may expect that the mutual inductance $M$ is a function of $H$. With the approach of the magnon–mode resonance, the microwave magnetic permeability of the YIG film [47] grows. This affects the microwave fields of the SRR. However, the microwave magnetic permeability in the vicinity of the feeding microstrip remains the same, since this part is not covered by the film. This applied–field dependent jump in the spatial profile of the permeability may make the strength inductive coupling of the microstrip to the SRR magnetic–field dependent.

## 4. Conclusions

By using both the frequency–domain VNA–FMR and time–domain PIMM spectroscopy techniques, we have demonstrated a strong coupling regime of magnons to microwave photons in the planar geometry of a lithographically formed split–ring resonator loaded by a single–crystal epitaxial YIG film. Whereas in the VNA–FMR experiment this interaction manifests itself as a strong anti–crossing between the photon and magnon mode, the time–domain PIMM traces exhibit a signature of a strong beat effect.

We have conducted numerical simulations of the microwave field structure of the SRR and of the magnetisation dynamics driven by the microwave currents in the SRR and suggested an equivalent circuit of an SRR loaded by a magnetic film. These calculations are in very good agreement with the experiment and reveal the physical origins of the effect of anti–crossing.

Our results are important for the progress of microwave quantum photonic devices such as, e.g., generators of entangled microwave radiation [48] required for quantum teleportation, quantum communication, or quantum radar with continuous variables at microwave frequencies [49]. Moreover, our findings are of immediate relevance to the development of nonlinear metamaterials

[50] and magnetically tuneable metamaterials [23] exploiting the strong coupling of magnons to microwave photons.

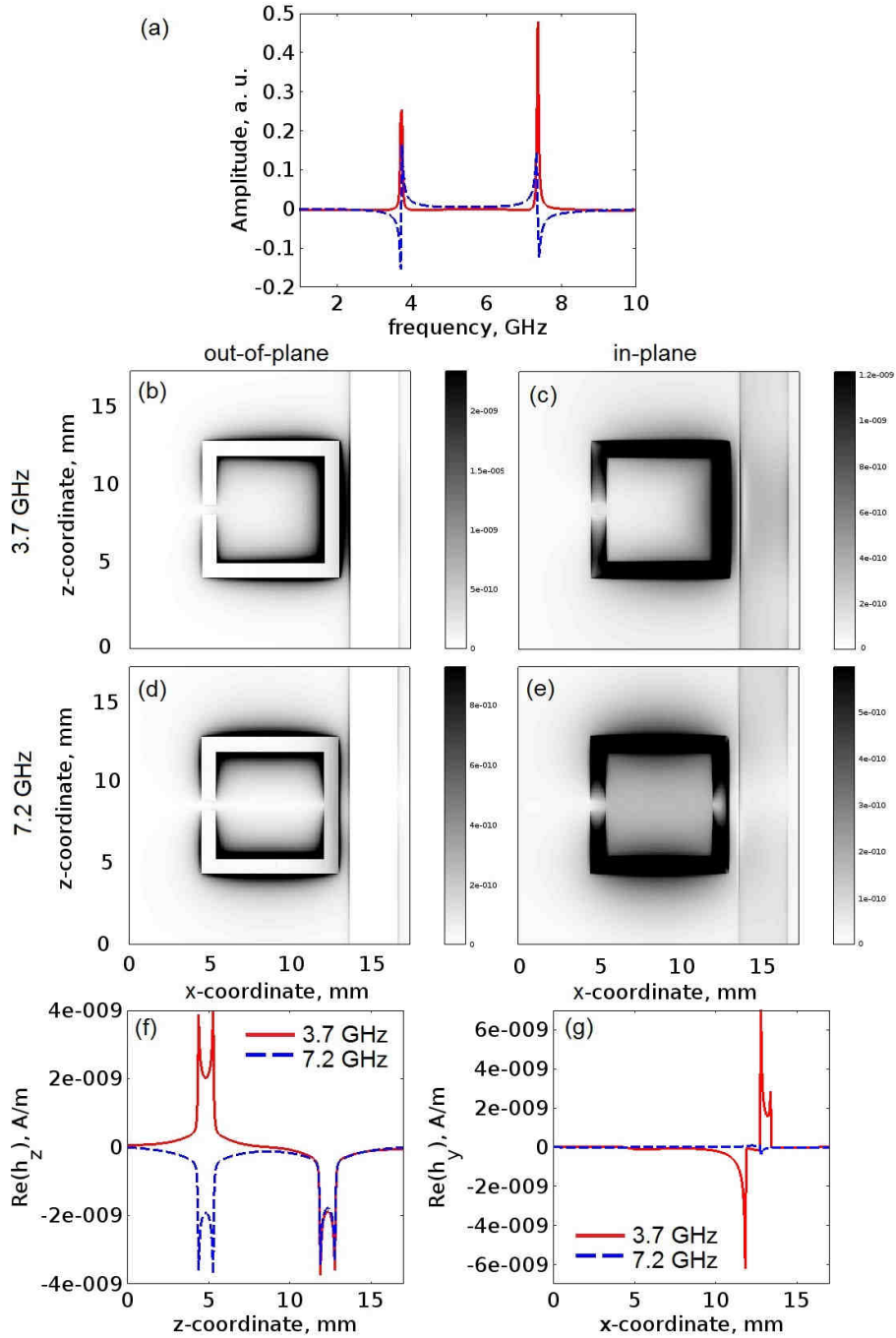

Fig. 6 Results of 3D FDTD simulations. The YIG film and its GGG substrate are not taken into account. (a) Spectrum of the SRR excited by the microstrip line. (b–e) Intensity profiles of the out–of–plane ($h_{out} = |h_y|$) and in–plane [$h_{in} = (|h_x|^2 + |h_z|^2)^{1/2}$] fields. The false colour is slightly oversaturated for the sake of illustration. (f, g) Re($h_z$)–field and Re($h_x$)–field profiles across the $z$–direction and $x$–direction, respectively, for the resonance frequencies 3.7 GHz (solid line) and 7.2 GHz (dashed line). These results, which are complementary to those in panels (b–e), show the phase relationship for the in–plane field components.

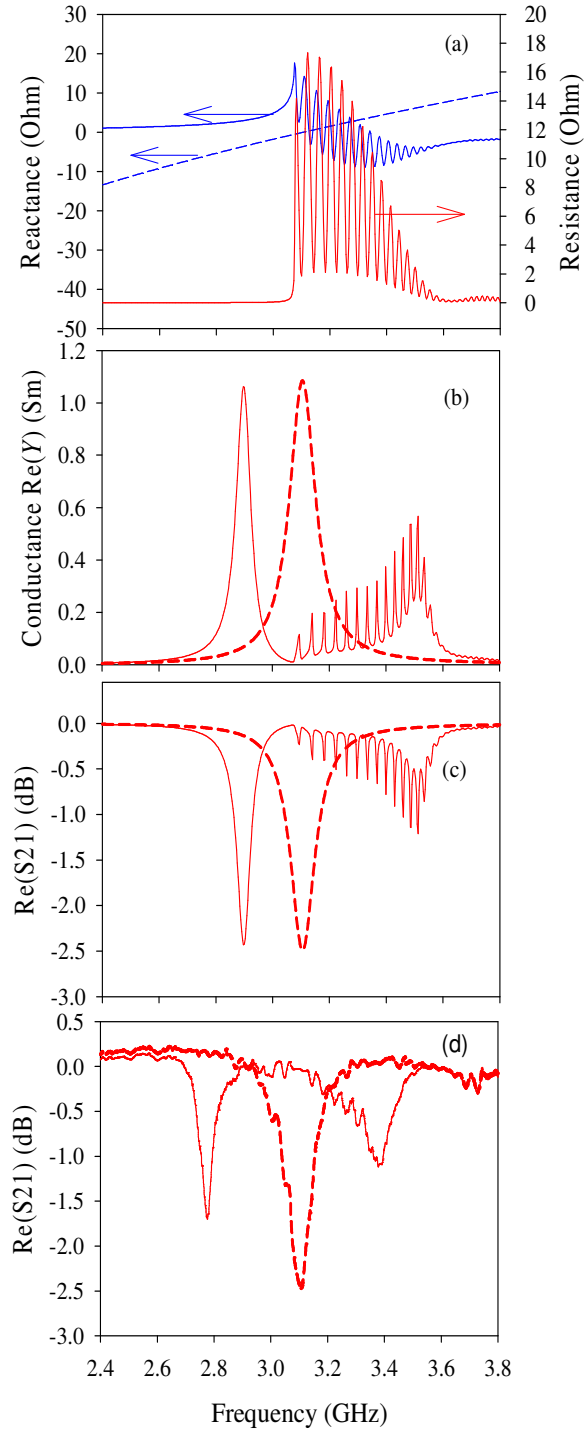

Fig. 7. (a) Complex radiation impedance $Z_r$ of two 0.5 mm–wide and 8.5 mm long parallel microstrip lines (solid lines). Left–hand axis: Im($Z_r$); right–hand axis: Re($Z_r$). The distance between the longitudinal axes of the microstrips is 8 mm. The static magnetic field is applied along the microstrips and is 485 Oe. Film saturation magnetisation $4\pi M_s = 2000$ G. This geometry mimics excitation of the Damon–Eshbach spin waves by the two sides of SRR which are parallel to the $x$–axis in Fig. 1. Dashed line: reactance for the uncoupled SRR (Im($X_0$)). (b) Calculated real part of conductance Re($Y$). Solid line: $H = 485$ Oe; dashed line: $H = 0$. (In the latter case it is assumed that $Z_r(H=0) = 0$). (c) Simulated signal from the output of the feeding microstrip line Re($S_{21}$). Solid line: $H=485$ Oe; dashed line: $H = 0$. (d) Experimental data for Re($S_{21}$). Solid line: $H = 470$ Oe; dashed line: $H = 0$.

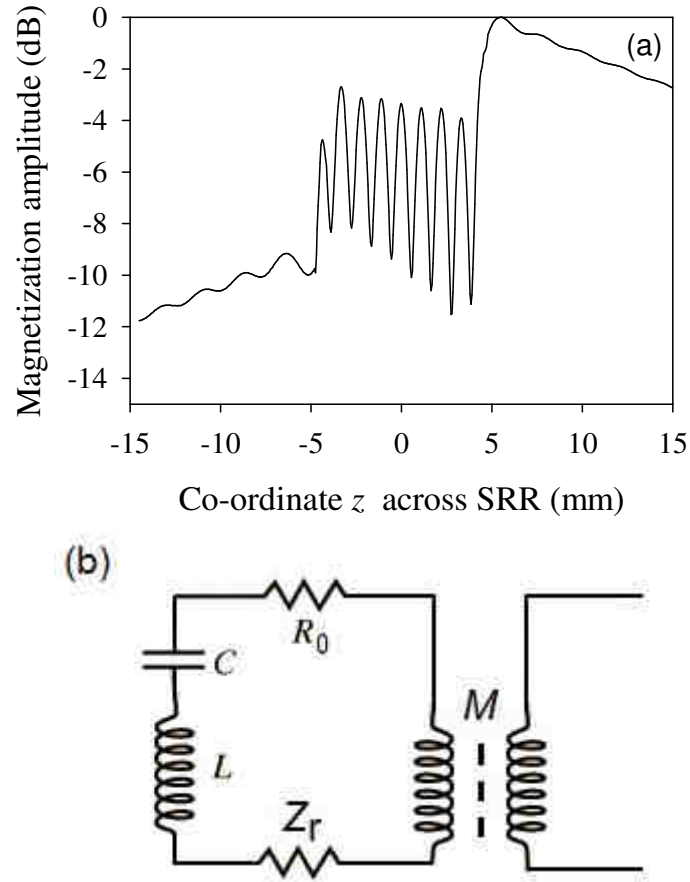

Fig. 8. (a) Calculated profile of the dynamic magnetisation amplitude in the z direction. The SRR sides are located at $z = \pm 4$ mm. Frequency is 3.242 GHz. The other parameters of calculation are the same as for Fig. 7. One notices a linear decay of the magnetisation amplitude with the distance from the SRR outside the ring. One also notices a standing–wave pattern inside the ring. Note that the slight waviness on top of the linear slopes is an artefact of the calculation. It originates from finite accuracy of the numerical inverse Fourier transform used to produce these data. (b) Equivalent circuit for the fundamental mode of SRR resonance loaded by the magnon mode in YIG.

**Acknowledgements**
Support by Department of Science and Technology of India, Australia–India Strategic Research Fund, and the Faculty of Science of the University of Western Australia is acknowledged. ISM has been supported by the UWA UPRF scheme.